\documentclass[prd,aps,showpacs,onecolumn]{revtex4}
\usepackage{amssymb}
\usepackage{amsfonts}
\usepackage{amsmath}
\usepackage{graphicx}
\usepackage{bm}

\input{tcilatex}
\begin{document}

\title{Exceptional orthogonal polynomials and generalized Schur polynomials }
\author{Yves Grandati}
\affiliation{LCP A2MC, Universit\'{e} de Lorraine, 1 Bd Arago, 57078 Metz, Cedex 3,
France.}

\begin{abstract}
We show that the exceptional orthogonal polynomials can be viewed as
confluent limits of the generalized Schur polynomials introduced by Sergeev
and Veselov.
\end{abstract}

\maketitle

\section{\protect\bigskip Introduction}

The concept of exceptional orthogonal polynomials (EOP) has been introduced
five years ago by Gomez-Ullate, Kamran and Milson \cite{gomez,gomez1}. They
form families of orthogonal polynomials whose degree sequences present a
finite number of gaps but which span a complete basis of their corresponding
Hilbert spaces. They appear to be closely related to exactly solvable
quantum systems built from translationally shape invariant potentials (TSIP)
via chains of Darboux-B\"{a}cklund transformations and this connection has
been the subject of an active research during the last years. The
exceptional Hermite, Laguerre and Jacobi polynomials are expressible as
Wronskians of the corresponding classical orthogonal polynomials and give,
up to a gauge factor the eigenstates of these extended potentials (see \cite%
{GGM1} and \cite{GGM} and references therein).

In this letter we show that these Wronskians can be viewed as confluent
limit of alternants which coincide with the generalized Schur polynomials
defined in their seminal paper by Sergeev and Veselov \cite{veselov} and
recently generalized by Harnad and Lee \cite{harnad}. The confluent limit of
the generalized Jacobi-Trudy formula established by Sergeev and Veselov is
also discussed.

\section{Chain of state-deleting Darboux-B\"{a}cklund transformations (SDDBT)%
}

\subsection{One step DBT}

We consider a one dimensional Hamiltonian $\widehat{H}=-d^{2}/dx^{2}+V(x),\
x\in I\subset \mathbb{R}$ and the associated Schr\"{o}dinger equation%
\begin{equation}
\psi _{\mu }^{\prime \prime }(x)+\left( E_{\mu }-V(x)\right) \psi _{\mu
}(x)=0,  \label{EdS}
\end{equation}%
$\psi _{\mu }(x)$ being a formal eigenfunction of $\widehat{H}$ for the
eigenvalue $E_{\mu }$. In the following we suppose that, with Dirichlet
boundary conditions on $I$, $\widehat{H}$ admits a discrete spectrum of
energies and eigenstates of the $\left( E_{n},\psi _{n}\right) _{n\in
\left\{ 0,...,n_{\max }\right\} \mathbb{\subseteq N}}$ where, without loss
of generality, we can always suppose that the ground level of $\widehat{H}$
is zero: $E_{0}=0$.

Starting from a given solution $\psi _{\nu }(x)$ associated to the value $%
\mu =\nu $ of the spectral parameter (eigenvalue $E_{\mu }$), we define the
first order operator $\widehat{A}\left( w_{\nu }\right) $ by

\begin{equation}
\widehat{A}\left( w_{\nu }\right) =d/dx+w_{\nu }(x),  \label{opA}
\end{equation}%
where $w_{\mu }(x)=-\psi _{\mu }^{\prime }(x)/\psi _{\mu }(x)$. For $\mu
\neq \nu $, the function defined via the Darboux-Crum formula 
\begin{equation}
\psi _{\mu }^{\left( \nu \right) }=\widehat{A}\left( w_{\nu }\right) \psi
_{\lambda }(x)=\frac{W\left( \psi _{\nu },\psi _{\mu }\mid x\right) }{\psi
_{\nu }(x)},  \label{DC}
\end{equation}%
where $W\left( y_{1},...,y_{m}\mid x\right) $ denotes the Wronskian of the
family of functions $y_{1},...,y_{m}$

\begin{equation}
W\left( y_{1},...,y_{m}\mid x\right) =\left\vert 
\begin{array}{ccc}
y_{1}\left( x\right) & ... & y_{m}\left( x\right) \\ 
... &  & ... \\ 
y_{1}^{\left( m-1\right) }\left( x\right) & ... & y_{m}^{\left( m-1\right)
}\left( x\right)%
\end{array}%
\right\vert ,  \label{wronskien}
\end{equation}%
is then a solution of the Schr\"{o}dinger equation%
\begin{equation}
\psi _{\mu }^{\left( \nu \right) \prime \prime }(x)+\left( E_{\mu
}-V^{\left( \nu \right) }(x)\right) \psi _{\mu }^{\left( \nu \right) }(x)=0,
\end{equation}%
with the same energy $E_{\lambda }$ as in Eq(\ref{EdS}) but with a modified
potential

\begin{equation}
V^{\left( \nu \right) }(x)=V(x)+2w_{\nu }^{\prime }(x),  \label{pottrans}
\end{equation}

We call $V^{\left( \nu \right) }(x)$ an extension of $V(x)$ and the
correspondence

\begin{equation}
\left( 
\begin{array}{c}
V(x) \\ 
\psi _{\mu }(x)%
\end{array}%
\right) \overset{A\left( w_{\nu }\right) }{\rightarrow }\left( 
\begin{array}{c}
V^{\left( \nu \right) }(x) \\ 
\psi _{\mu }^{\left( \nu \right) }(x)%
\end{array}%
\right) 
\end{equation}%
is called a Darboux-B\"{a}cklund Transformations (DBT). The eigenfunction $%
\psi _{\nu }$ is the seed function of the DBT $A(w_{\nu })$.

Note that $\widehat{A}(w_{\nu })$ annihilates $\psi _{\nu }$ and
consequently the formula Eq(\ref{DC}) allows to obtain an eigenfunction of $%
V^{\left( \nu \right) }$ for the eigenvalue $E_{\mu }$ only when $\mu \neq
\nu $. Nevertheless, we can readily verify that $1/\psi _{\nu }(x)$ is such
an eigenfunction. By extension, we then define the "image" by $A(w_{\nu })$
of the seed eigenfunction $\psi _{\nu }$ itself as 
\begin{equation}
\psi _{\nu }^{\left( \nu \right) }(x)\sim 1/\psi _{\nu }(x).  \label{psinunu}
\end{equation}

\subsection{Formal chains of DBT}

At the formal level, the DBT can be straightforwardly iterated and a chain
of $m$\ DBT is simply described by the following scheme

\begin{equation}
\left\{ 
\begin{array}{c}
\psi _{\mu }\overset{A(w_{\nu _{1}})}{\rightarrowtail }\psi _{\mu }^{\left(
\nu _{1}\right) }\overset{A(w_{\nu _{2}}^{\left( N_{1}\right) })}{%
\rightarrowtail }\psi _{\mu }^{\left( N_{2}\right) }...\overset{A(w_{\nu
_{m}}^{\left( N_{m-1}\right) })}{\rightarrowtail }\psi _{\mu }^{\left(
N_{m}\right) } \\ 
V\overset{A(w_{\nu _{1}})}{\rightarrowtail }V^{\left( \nu _{1}\right) }%
\overset{A(w_{\nu _{2}}^{\left( N_{1}\right) })}{\rightarrowtail }V^{\left(
N_{2}\right) }...\overset{A(w_{\nu _{m}}^{\left( N_{m-1}\right) })}{%
\rightarrowtail }V^{\left( N_{m}\right) },%
\end{array}%
\right.   \label{diagn}
\end{equation}%
where $N_{j}$ denotes the $j$-uple $\left( \nu _{1},...,\nu _{j}\right) $
(with $N_{1}=\nu _{1}$) of\textbf{\ }spectral indices which completely
characterizes the chain. We note $\left( N_{m},\nu _{m+1},...,\nu
_{m+k}\right) $ the chain obtained by adding to the chain $N_{m}$ the DBT
associated to the successive eigenfunctions $\psi _{\nu _{m+1}}^{\left(
N_{m}\right) },...,\psi _{\nu _{m+k}}^{\left( N_{m+k-1}\right) }$.

$\psi _{\mu }^{\left( N_{m}\right) }$ is an eigenfunction associated to the
eigenvalue $E_{\mu }$ of the potential (see Eq(\ref{pottrans}))

\begin{equation}
V^{\left( N_{m}\right) }(x)=V(x)+2\sum_{j=1}^{m}\left( w_{\nu _{j}}^{\left(
N_{j-1}\right) }(x)\right) ^{\prime }=V(x)-2\sum_{j=1}^{m}\left( \log \left(
\psi _{\nu _{j}}^{\left( N_{j-1}\right) }(x)\right) \right) ^{\prime \prime
}.  \label{potnstep}
\end{equation}

It can be written as (cf Eq(\ref{DC}))

\begin{equation}
\psi _{\mu }^{\left( N_{m}\right) }(x)=\widehat{A}\left( w_{\nu
_{m}}^{\left( N_{m-1}\right) }\right) \psi _{\mu }^{\left( N_{m-1}\right)
}(x)=\widehat{A}\left( w_{\nu _{m}}^{\left( N_{m-1}\right) }\right) ...%
\widehat{A}\left( w_{\nu _{1}}\right) \psi _{\mu }(x),  \label{etats n}
\end{equation}%
that is,

\begin{equation}
\psi _{\mu }^{\left( N_{m}\right) }(x)=\frac{W\left( \psi _{\nu
_{m}}^{\left( N_{m-1}\right) },\psi _{\mu }^{\left( N_{m-1}\right) }\mid
x\right) }{\psi _{\nu _{m}}^{\left( N_{m-1}\right) }(x)}.  \label{fon}
\end{equation}

A chain is non-degenerate if all the spectral indices $\nu _{i}$ of the
chain $N_{m}$ are distinct and is degenerate if some of them are repeated in
the chain. For non-degenerate chains, Crum has established formulas which
give the extended potentials and their eigenfunctions in terms of Wronskians
of eigenfunctions of the initial potential \cite{crum,matveev,GGM1}.

\textbf{Crum's formulas}

\textit{When all the }$\nu _{j}$\textit{\ and }$\lambda $\textit{\ are
distinct, we have}

\begin{equation}
\psi _{\mu }^{\left( N_{m}\right) }(x)=\frac{W^{\left( N_{m},\mu \right)
}\left( x\right) }{W^{\left( N_{m}\right) }\left( x\right) }
\label{etats n3}
\end{equation}%
\textit{and}

\begin{equation}
V^{\left( N_{m}\right) }(x)=V(x)-2\left( \log W^{\left( N_{m}\right) }\left(
x\right) \right) ^{\prime \prime },  \label{potnstep2}
\end{equation}%
\textit{where }$W^{\left( N_{m}\right) }\left( x\right) =W\left( \psi _{\nu
_{1}},...,\psi _{\nu _{m}}\mid x\right) $.

The eigenfunctions $\psi _{\nu _{1}},...,\psi _{\nu _{m}}$ of $V$ are called
the seed functions of the chain of DBT associated to the $m$-uple of
spectral indices $N_{m}=\left( \nu _{1},...,\nu _{m}\right) $.$\ \ \ \ $

\subsection{Chains of SDDBT and partitions}

In the following, we call state-deleting DBT (SDDBT) every DBT whose seed
function is an eigenstate. A chain of SDDBT is then characterized by an\ $m$%
-uple $N_{m}=\left( n_{1},,...,n_{m}\right) ,\ n_{i+1}>n_{i}\geq 0,$\ of
distinct positive integers such that the corresponding seed functions $\psi
_{n_{i}}$\ are eigenstates of the initial potential.

Consider a chain of SDDBT associated to a $m$-uple of spectral indices $%
N_{m}=\left( n_{1},...,n_{m}\right) $ with $n_{m}>...>n_{1}\geq 0$. We can
equivalently characterize this chain by a partition $\lambda =\left( \lambda
_{1},...,\lambda _{m}\right) $ of length $l(\lambda )=m$, where $\lambda
_{1}>...>\lambda _{m}\geq 0$ and

\begin{equation}
\lambda _{i}=n_{m-i+1}-m+i.  \label{lambda-n}
\end{equation}

$\lambda $ is a partition of the integer

\begin{equation}
\left\vert \lambda \right\vert =\sum_{i=1}^{m}\lambda
_{i}=\sum_{i=1}^{m}n_{i}-m(m-1)/2.
\end{equation}

Note that, contrarily to the usual convention, we authorize $\lambda $ to
contain at its end a string of zeros (this string corresponding to a
complete chain of SDDBT). The reduced form\textbf{\ }$\widetilde{\lambda }$
of the partition is obtained by suppressing this last string of zeros. If $%
\lambda $ does not contain such a chain we say that it is an irreducible
partition.

To this partition is associated a Young diagram which characterizes the
spectral shape of the extension. The $\lambda _{i}$ are the lengths of the
columns of the Young diagram starting from the left corner. Following \cite%
{veselov1,GGM}, we define the double partition $\lambda ^{2}$ of $\lambda $
as the partition of length $l\left( \lambda ^{2}\right) =2m$ defined as $%
\lambda ^{2}=\left( \lambda _{1}^{2},...,\lambda _{m}^{2}\right) $, where we
note $\lambda _{i}^{k}$ for $\lambda _{i}$ repeated $k$ times ($\lambda
_{i}^{k}=$ $\lambda _{i},...,\lambda _{i}$). We call an Adler partition a
partition $\lambda $ whose reduced form, $\widetilde{\lambda }=\left(
\lambda _{1}^{m_{1}},...,\lambda _{k}^{m_{k}}\right) $, is a double
partition.

The qualifier state-deleting used above is somewhat abusive, since strictly
speaking it have to be reserved to chains leading to regular potentials.
Indeed, in this case, the action on the spectrum of the successive DBT based
on eigenstates corresponds to suppress at each step the level associated to
the used seed function, which then justifies the denomination
"state-deleting" DBT.

Krein \cite{krein} and later Adler \cite{adler} have given a necessary and
sufficient regularity condition for the final extensions of such non
degenerate chains of SDDBT. The Krein-Adler theorem can be rewritten as a
structural condition for the partition associated to the chain of SDDBT \cite%
{krein,adler,GGM} as follows

\textbf{Krein-Adler theorem}

\textit{The final extension of the chain of SDDBT associated to the \ }$m$%
\textit{-uple }$N_{m}=\left( n_{1},,...,n_{m}\right) ,\ n_{i+1}>n_{i}\geq 0,$%
\textit{\ or equivalently to the partition }$\lambda =\left( \lambda
_{1}^{m_{1}},...,\lambda _{k}^{m_{k}},0^{r}\right) ,\
\dsum\limits_{i=1}^{k}m_{k}+r=m,$\textit{\ is regular iff }$\lambda $\textit{%
\ is an Adler partition, that is, iff }$m_{i}\in 2\mathbb{N},\ \forall i\in
\left\{ 1,...,k\right\} $\textit{. The spectrum of the final extension }$%
V^{\left( N_{m}\right) }(x)$\textit{\ contains only }even gaps\textit{\
(gaps constituted by an even number of consecutive missing levels).}

\textit{Such a chain of SDDBT is said of the }Krein-Adler type\textit{\ and
in the particular case where }$\widetilde{\lambda }=0$\textit{, the chain is
said to be \textbf{complete}.}

\section{Jacobi-Trudi formula for EOP}

\subsection{PTSIP}

Consider a potential $V(x;a)$ which depends upon a (multi)parameter $a\in 
\mathbb{R}^{N}$ and which admits a (finite or infinite) bound state spectrum 
$\left( E_{n},\psi _{n}\right) _{n\geq 0}$, the ground level being supposed
to be zero: $E_{0}(a)=0$. In the framework of SUSY QM, such a potential\ is
said to be shape invariant (SIP) \cite{cooper,Dutt,gendenshtein} if its SUSY
partner

\begin{equation}
V^{\left( 0\right) }(x;a)=V(x;a)+2w_{0}^{\prime }(x;a),
\end{equation}%
keeps the same functional form as the initial potential. Namely

\begin{equation}
V^{\left( 0\right) }(x;a)=V(x;f(\alpha ))+R(a),  \label{SI}
\end{equation}%
$R\left( a\right) \in \mathbb{R}$ and $f(a)\in \mathbb{R}^{N}$ being two
given functions of $a$.

In this case, it can be shown\ \cite{cooper,Dutt,gendenshtein} that the
complete bound state energy spectrum of $\widehat{H}(a)=-\frac{d^{2}}{dx^{2}}%
+V(x;a)$ is given by:

\begin{equation}
E_{n}(a)=\sum_{k=0}^{n-1}R(a_{k})=\sum_{i=0}^{n-1}E_{1}(a_{i}),
\label{spectreSIP}
\end{equation}%
where $a_{k}=f^{\left( k\right) }(a)=\overset{\text{k times}}{\overbrace{%
f\circ ...\circ f}}(a)$.

As for the corresponding eigenstates, they can be written as

\begin{equation}
\psi _{n}(x;a)\sim \widehat{A}^{+}(a)\psi _{n-1}(x;a_{1})\sim \widehat{A}%
^{+}(a)...\widehat{A}^{+}(a_{n-1})\psi _{0}(x;a_{n}),  \label{spectreSIP2}
\end{equation}%
where $\widehat{A}^{+}(a)=-\frac{d}{dx}+w_{0}(x;a)$.

When $f$ is a simple translation $f(a)=a+\varepsilon ,\ \varepsilon =\left(
\varepsilon ^{\left( 1\right) },...,\varepsilon ^{\left( N\right) }\right)
\in \mathbb{R}^{N}$, $V$ is said to be translationally shape invariant and
we call it a TSIP. For all the known TSIP we have $a\in 
\mathbb{R}
$ (first category TSIP) or $a\in 
\mathbb{R}
^{2}$ (second category TSIP) \cite{cooper,Dutt,grandati}.

The set of TSIP contains all the potentials classicaly known to be exactly
solvable, ie for which we know explicitely the dispersion relation and whose
the eigenfunctions can be expressed in closed analytical form in terms of
elementary transcendental functions: the harmonic, isotonic, Morse,
Kepler-Coulomb, Eckart, Darboux-P\"{o}schl-Teller (hyperbolic and
trigonometric) and Rosen-Morse (hyperbolic and trigonometric) potentials.
These potentials are primary TSIP (PTSIP) from which it is possible in some
cases to build infinite towers of secondary TSIP (STSIP) which are
extensions of the previous ones and which share the same translational shape
invariance properties \cite{grandati5}.

An important feature of the the PTSIP is that their eigenfunctions $\psi _{n}
$ are equal, up to a gauge factor, to classical orthogonal polynomials in an
appropriate variable $z$ (which can be $n$ dependent) and we say that $\psi
_{n}$ is quasi-polynomial in this variable. The confining (ie diverging at
both boundaries of the definition intrerval) PTSIP, which then possess an
infinite bound state spectrum, are the harmonic, isotonic and trigonometric
Darboux-P\"{o}schl-Teller (TDPT) potentials. For these ones, the gauge
factor and the adapted variable are independent of $n$ and their
(unnormalized) eigenstates can be written as

\begin{equation}
\psi _{n}(x;a)=\psi _{0}(x;a)\Pi _{n}^{a}(z\left( x\right) ),
\label{eigenTSIP}
\end{equation}%
where $\Pi _{n}^{a}(z)$ is a monic classical orthogonal polynomial (Hermite,
Laguerre and Jacobi) \cite{magnus,szego}.

The $\Pi _{n}^{a}\left( z\right) $ satisfy the recursion relation

\begin{equation}
\left\{ 
\begin{array}{c}
\Pi _{0}^{a}(z)=1 \\ 
z\Pi _{n}^{a}(z)=\Pi _{n+1}^{a}(z)+p_{n,a}\Pi _{n}^{a}(z)+q_{n,a}\Pi
_{n-1}^{a}(z)%
\end{array}%
\right. ,  \label{recphi}
\end{equation}%
with

\begin{equation}
\left\{ 
\begin{array}{c}
p_{n,a}=0 \\ 
q_{n,a}=q_{n}=n/2.%
\end{array}%
\right. ,\text{ for the Hermite case,}  \label{recH}
\end{equation}

\begin{equation}
\left\{ 
\begin{array}{c}
p_{n,a}=p_{n,\alpha }=2n+1+\alpha \\ 
q_{n,a}=q_{n,\alpha }=n(n+\alpha ).%
\end{array}%
\right. ,\text{ for the Laguerre case,}  \label{recL}
\end{equation}%
and

\begin{equation}
\left\{ 
\begin{array}{c}
p_{n,a}=p_{n,\alpha ,\beta }=\frac{\beta ^{2}-\alpha ^{2}}{\left(
2n+2+\alpha +\beta \right) \left( 2n+\alpha +\beta \right) } \\ 
q_{n,a}=q_{n,\alpha ,\beta }=\frac{4n(n+\alpha )(n+\beta )(n+\alpha +\beta )%
}{\left( 2n+1+\alpha +\beta \right) \left( 2n+\alpha +\beta \right)
^{2}\left( 2n-1+\alpha +\beta \right) }.%
\end{array}%
\right. ,\text{ for the Jacobi case.}  \label{recJ}
\end{equation}

In the following, we give a resumed description of the spectral properties
of the three confining PTSIP.

\subsubsection{Harmonic oscillator}

The harmonic oscillator (HO) potential (with zero ground level $E_{0}=0$))
is defined on the real line by

\begin{equation}
V\left( x;\omega \right) =\frac{\omega ^{2}}{4}x^{2}-\frac{\omega }{2},\
\omega \in 
\mathbb{R}
^{+}.  \label{OH}
\end{equation}

With Dirichlet boundary conditions at infinity it has the following spectrum
($z(x)=\sqrt{\omega /2}x$)%
\begin{equation}
\left\{ 
\begin{array}{c}
E_{n}\left( \omega \right) =n\omega \\ 
\psi _{n}(x;\omega )=\psi _{0}(x;\omega )\widetilde{H}_{n}\left( z\right)%
\end{array}%
\right. ,\quad n\geq 0,  \label{spec OH}
\end{equation}%
with

\begin{equation}
\psi _{0}(x;\omega )=\exp \left( -z^{2}/2\right)  \label{psi0HO}
\end{equation}%
and

\begin{equation}
\widetilde{H}_{n}\left( z\right) =\frac{1}{2^{n}}H_{n}\left( z\right) ,
\end{equation}%
the $H_{n}\left( z\right) $ being the classical Hermite polynomials.

It is the most simple example of TSIP, with $a=\omega \in 
\mathbb{R}
$ and $\varepsilon =0$ (the parameter translation is of zero amplitude $%
a_{1}=\omega $), that is

\begin{equation}
V^{\left( 0\right) }\left( x;\omega \right) =V\left( x;\omega \right)
+\omega .
\end{equation}

\subsubsection{Isotonic oscillator}

The isotonic oscillator (IO) potential (with zero ground level $E_{0}=0$))
is defined on the positive half line $\left] 0,+\infty \right[ $ by

\begin{equation}
V\left( x;\omega ,\alpha \right) =\frac{\omega ^{2}}{4}x^{2}+\frac{\left(
\alpha +1/2\right) (\alpha -1/2)}{x^{2}}-\omega \left( \alpha +1\right)
,\quad \alpha >1/2.  \label{OI}
\end{equation}

If we add Dirichlet boundary conditions at $0$ and infinity and if we
suppose $\alpha >1/2$, it has the following spectrum ($z(x)=\omega x^{2}/2$)%
\begin{equation}
\left\{ 
\begin{array}{c}
E_{n}\left( \omega \right) =2n\omega \\ 
\psi _{n}\left( x;\omega ,\alpha \right) =\psi _{0}\left( x;\omega ,\alpha
\right) \widetilde{L}_{n}^{\alpha }\left( z\right)%
\end{array}%
\right. ,\quad n\geq 0,  \label{spec OI}
\end{equation}%
where

\begin{equation}
\psi _{0}\left( x;\omega ,\alpha \right) =z^{\left( \alpha +1/2\right)
/2}e^{-z/2}  \label{psi0IO}
\end{equation}%
and

\begin{equation}
\widetilde{L}_{n}^{\alpha }\left( z\right) =\left( -1\right)
^{n}n!L_{n}^{\alpha }\left( z\right) ,
\end{equation}%
\ the $L_{n}^{\alpha }\left( z\right) \ $being the classical Laguerre
polynomials.

It is a TSIP, with $a=\left( \omega ,\alpha \right) \in 
\mathbb{R}
^{2}$ and $\varepsilon =\left( 0,+1\right) $

\begin{equation}
V^{\left( 0\right) }\left( x;\omega ,\alpha \right) =V\left( x;\omega
,\alpha _{1}\right) +2\omega .
\end{equation}

\subsubsection{TDPT}

The trigonometric Darboux-P\"{o}schl-Teller (TDPT) potential is defined on
the interval $\left] 0,\pi /2\right[ $ by

\begin{equation}
V(x;\alpha ,\beta )=\frac{(\alpha +1/2)(\alpha -1/2)}{\cos ^{2}x}+\frac{%
(\beta +1/2)(\beta -1/2)}{\sin ^{2}x}-(\alpha +\beta +1)^{2},\ \alpha ,\beta
>1/2.  \label{TDPT}
\end{equation}

With Dirichlet boundary conditions at $0$ and $\pi /2$, it has the following
spectrum 
\begin{equation}
\left\{ 
\begin{array}{c}
\text{ }E_{n}\left( \alpha ,\beta \right) =(\alpha _{n}+\beta
_{n}+1)^{2}-(\alpha +\beta +1)^{2}=4n(\alpha +\beta +1+n) \\ 
\\ 
\psi _{n}\left( x;\alpha ,\beta \right) =\psi _{0}\left( x;\alpha ,\beta
\right) \widetilde{P}_{n}^{\left( \alpha ,\beta \right) }\left( \cos
2x\right)%
\end{array}%
\right. ,\ n\in \mathbb{N},  \label{spec TDPT}
\end{equation}%
where

\begin{equation}
\psi _{0}\left( x;\alpha ,\beta \right) =\left( \sin x\right) ^{\alpha
+1/2}\left( \cos x\right) ^{\beta +1/2}  \label{psi0TDPT}
\end{equation}%
and

\begin{equation}
\widetilde{P}_{n}^{\left( \alpha ,\beta \right) }\left( z\right) =\frac{%
2^{n}n!}{\left( \alpha +\beta +n+1\right) _{n}}P_{n}^{\left( \alpha ,\beta
\right) }\left( z\right) .
\end{equation}

The $\mathit{P}_{n}^{\left( \alpha ,\beta \right) }$ are the usual Jacobi
polynomials, $\left( x\right) _{n}=x(x+1)...(x+n-1)\ $is the Pochhammer
symbol \cite{szego,magnus} and $(\alpha _{n},\beta _{n})=(\alpha +n,\beta
+n) $.

It is a TSIP, with $a=\left( \alpha ,\beta \right) \in 
\mathbb{R}
^{2}$ and $\varepsilon =\left( +1,+1\right) $

\begin{equation}
V^{\left( 0\right) }\left( x;\alpha ,\beta \right) =V\left( x;\alpha
_{1},\beta _{1}\right) +4(\alpha +\beta +2).
\end{equation}

\subsection{Jacobi-Trudi type formula for EOP}

Due to the Crum formulas Eq(\ref{etats n3}) and Eq(\ref{potnstep2}), the
form of the extensions obtained from a PTSIP via chains of SDDBT as well as
their eigenfunctions are determined by Wronskians of the type

\begin{equation}
W^{\left( N_{m}\right) }\left( x;a\right) =W\left( \psi _{n_{1}}\left(
x;a\right) ,...,\psi _{n_{m}}\left( x;a\right) \mid x\right) ,  \label{W1}
\end{equation}%
where the $\psi _{n}$ are given by Eq(\ref{eigenTSIP}). $W^{\left(
N_{m}\right) }$ is characterized by the m-uple of spectral indices $%
N_{m}=\left( n_{1},...,n_{m}\right) $ ($n_{m}>...>n_{1}\geq 0$) or
equivalently by the associated partition $\lambda =\left( \lambda
_{1},...,\lambda _{m}\right) $ (see Eq(\ref{lambda-n})). Considering the
three confining PTSIP mentioned above, using the standard properties of
Wronskians \cite{muir}, this can be rewritten

\begin{equation}
W^{\left( N_{m}\right) }\left( x;a\right) =\left( \psi _{0}(x;a)\right)
^{m}W\left( \Pi _{n_{1}}^{a}(z),...,\Pi _{n_{m}}^{a}(z)\mid x\right) =\left(
\psi _{0}(x;a)\right) ^{m}\left( \frac{dz}{dx}\right) ^{m(m+1)/2}\mathcal{W}%
_{\lambda }\left( z\right) ,  \label{W3}
\end{equation}%
where (see Eq(\ref{lambda-n}))

\begin{equation}
\mathcal{W}_{\lambda }\left( z\right) =W\left( \Pi _{n_{1}}^{a}(z),...,\Pi
_{n_{m}}^{a}(z)\mid z\right) =W\left( \Pi _{\lambda _{m}}^{a}(z),...,\Pi
_{\lambda _{1}+m-1}^{a}(z)\mid z\right)   \label{EOP}
\end{equation}%
is a polynomial in $z$ that we call in an abusive manner an exceptional
orthogonal polynomial (EOP) enlarging the denomination associated to regular
extensions \cite{gomez}. Since, as we have seen above, the regularity is
only related to a particular structure of the partition $\lambda $, we don't
refer to it in the following.

For the monic classical orthogonal polynomials, we have \cite{szego,magnus}

\begin{equation}
\frac{d}{dz}\Pi _{n}^{a}(z)=n\Pi _{n-1}^{a_{1}}(z),  \label{derivmonic}
\end{equation}%
which gives

\begin{equation}
\mathcal{W}_{\lambda }\left( z\right) =\left\vert 
\begin{array}{cccc}
A_{\lambda _{m}}^{0}\Pi _{\lambda _{m}}^{a}(z) & A_{\lambda _{m-1}+1}^{0}\Pi
_{\lambda _{m-1}+1}^{a}(z) & ... & A_{\lambda _{1}+m-1}^{0}\Pi _{\lambda
_{1}+m-1}^{a}(z) \\ 
A_{\lambda _{m}}^{1}\Pi _{\lambda _{m}-1}^{a_{1}}(z) & A_{\lambda
_{m-1}+1}^{1}\Pi _{\lambda _{m-1}}^{a_{1}}(z) & ... & A_{\lambda
_{1}+m-1}^{1}\Pi _{\lambda _{1}+m-2}^{a_{1}}(z) \\ 
... & ... &  & ... \\ 
A_{\lambda _{m}}^{m-1}\Pi _{\lambda _{m}-m+1}^{a_{m-1}}(z) & A_{\lambda
_{m-1}+1}^{m-1}\Pi _{\lambda _{m-1}-m+2}^{a_{m-1}}(z) & ... & A_{\lambda
_{1}+m-1}^{m-1}\Pi _{\lambda _{1}}^{a_{m-1}}(z)%
\end{array}%
\right\vert ,  \label{EOPdet}
\end{equation}%
where

\begin{equation}
A_{n}^{k}=n(n-1)...(n-k+1)=\frac{n!}{(n-k)!}.  \label{arrangement}
\end{equation}

Defining

\begin{equation}
\mathit{g}_{k}^{\left( j\right) }\left( z\right) =A_{k+m-1-j}^{m-1-j}\Pi
_{k}^{a_{m-1}+j}(z)=\frac{\left( k-1+m-j\right) !}{k!}\Pi _{k}^{a_{m-1}+j}(z)
\label{g}
\end{equation}
and reversing the order of the columns and of the lines, we arrive to%
\begin{equation}
\mathcal{W}_{\lambda }\left( z\right) =\left\vert 
\begin{array}{cccc}
\mathit{g}_{\lambda _{1}}^{\left( 0\right) }\left( z\right) & \mathit{g}%
_{\lambda _{2}-1}^{\left( 0\right) }\left( z\right) & ... & \mathit{g}%
_{\lambda _{m}-m+1}^{\left( 0\right) }\left( z\right) \mathit{\ } \\ 
\mathit{g}_{\lambda _{1}+1}^{\left( 1\right) }\left( z\right) & \mathit{g}%
_{\lambda _{2}}^{\left( 1\right) }\left( z\right) & ... & \mathit{g}%
_{\lambda _{m}-m+2}^{\left( 1\right) }\left( z\right) \\ 
... & ... &  & ... \\ 
\mathit{g}_{\lambda _{1}+m-1}^{\left( m-1\right) }\left( z\right) & \mathit{g%
}_{\lambda _{2}+m-2}^{\left( m-1\right) }\left( z\right) & ... & \mathit{g}%
_{\lambda _{m}}^{\left( m-1\right) }\left( z\right)%
\end{array}%
\right\vert .  \label{NoumiJT}
\end{equation}

Consequently, the EOP $\mathcal{W}_{\lambda }\left( z\right) $ is amenable
of a\textbf{\ }Jacobi-Trudi type formula\textbf{\ }a la Noumi \cite{noumi},
analogous to the one satisfied by the $\phi $-factors in Noumi-Yamada
approach of the rational solutions of the Painlev\'{e} equations. It has to
be noticed that the generalized Hermite and Okamoto polynomials \cite%
{clarkson,clarkson1} appear as particular exceptional Hermite polynomials in
the enlarged sense given above.

\section{Wronskians and confluent alternants}

Consider the following alternant \textbf{\ }\cite{muir}

\begin{equation}
\Delta \left( \Phi \mid X\right) =\left\vert 
\begin{array}{ccc}
\phi _{1}\left( x_{1}\right) & ... & \phi _{1}\left( x_{m}\right) \\ 
... &  & ... \\ 
\phi _{m}\left( x_{1}\right) & ... & \phi _{m}\left( x_{m}\right)%
\end{array}%
\right\vert ,  \label{alt}
\end{equation}%
where the $\phi _{i}$ are supposed polynomials of degree $n_{i}$ and where
we have noted $\Phi =\left( \phi _{1},...,\phi _{m}\right) $ and $X=\left(
x_{1},...,x_{m}\right) $. For the case $\phi _{k}\left( x\right) =x^{k-1}$
the preceding determinant reduces to a Vandermondian

\begin{equation}
\Delta \left( X\right) =\left\vert 
\begin{array}{ccc}
1 & ... & 1 \\ 
x_{1} & ... & x_{m} \\ 
... &  & ... \\ 
x_{1}^{m-1} & ... & x_{m}^{m-1}%
\end{array}%
\right\vert =\dprod\limits_{1\leq i<j\leq m}\left( x_{i}-x_{j}\right) .
\label{vand}
\end{equation}

$\Delta \left( \Phi \mid X\right) $ being a polynomial antisymmetric in the
exchange of two variables $x_{i}$ and $x_{j}$ is divisible $\Delta \left(
X\right) $ and the ratio $\frac{\Delta \left( \Phi \mid X\right) }{\Delta
\left( X\right) }$ is a symmetric polynomial $S\left( \Phi \mid X\right) $.

In the case where the $\phi _{k}$ are monic polynomials of respective
degrees $k,$ then by linear combinations of the columns, we obtain

\begin{equation}
\Delta \left( \Phi \mid X\right) =\Delta \left( X\right) .
\label{Deltamonic}
\end{equation}

We are interested in the confluent limit $x_{i}\rightarrow x,\ \forall i\in
\left\{ 1,...,m\right\} $. Defining the new set of variable $\varepsilon _{k}
$ via $x_{k}=x+\varepsilon _{k}$, we have

\begin{equation}
\phi _{i}\left( x+\varepsilon _{j}\right) =\sum_{k=0}^{n_{i}}a_{k,i}\left(
x\right) \varepsilon _{j}^{k}=p_{i}\left( \varepsilon _{j}\right) ,
\label{phi1}
\end{equation}%
with $a_{k,i}\left( x\right) =\phi _{i}^{\left( k\right) }\left( x\right)
/k! $ and

\begin{equation}
\Delta \left( \Phi \mid X\right) =\left\vert 
\begin{array}{ccc}
p_{1}\left( \varepsilon _{1}\right) & ... & p_{1}\left( \varepsilon
_{m}\right) \\ 
... &  & ... \\ 
p_{m}\left( \varepsilon _{1}\right) & ... & p_{m}\left( \varepsilon
_{m}\right)%
\end{array}%
\right\vert =\Delta \left( P\mid \varepsilon \right) ,  \label{deltaeps}
\end{equation}%
where $P=\left( p_{1},...,p_{m}\right) $ and $\varepsilon =\left(
\varepsilon _{1},...,\varepsilon _{m}\right) $. Moreover

\begin{equation}
\Delta \left( X\right) =\dprod\limits_{1\leq i<j\leq m}\left( \varepsilon
_{i}-\varepsilon _{j}\right) =\Delta \left( \varepsilon \right) ,
\label{vandeps}
\end{equation}%
which implies

\begin{equation}
S\left( \Phi \mid X\right) =S\left( P\mid \varepsilon \right)  \label{S}
\end{equation}

The ratio $S\left( P\mid \varepsilon \right) =\Delta \left( P\mid
\varepsilon \right) /\Delta \left( \varepsilon \right) $ is also a symmetric
polynomial and is consequently a continuous function of $\varepsilon $ on $%
\mathbb{R}^{m}$. It results in particular that to calculate the value $%
S\left( P\mid 0\right) $, we can successively apply the limits $\varepsilon
_{1}\rightarrow 0,\ \varepsilon _{2}\rightarrow 0,...,\ \varepsilon
_{m}\rightarrow 0$ in this order. We then have the following result

\textbf{Theorem 1:}

\textit{In the confluent limit }$x_{i}\rightarrow x,\ \forall i\in \left\{
1,...,m\right\} $

\begin{equation}
S\left( \Phi \mid X\right) =\frac{\Delta \left( \Phi \mid X\right) }{\Delta
\left( X\right) }\underset{\left\{ x_{i}\rightarrow x\right\} }{\rightarrow }%
\frac{W\left( \phi _{1},...,\phi _{m}\mid x\right) }{\dprod%
\limits_{j=1}^{m-1}j!}.
\end{equation}

\textbf{Proof:}

From Eq(\ref{S}) and Eq(\ref{phi1}), we have

\begin{equation}
S\left( \Phi \mid X\right) =S\left( P\mid \varepsilon \right) =\frac{1}{%
\Delta (\varepsilon )}\left\vert 
\begin{array}{ccc}
\dsum\limits_{k=0}^{n_{1}}a_{k,1}\left( x\right) \varepsilon _{1}^{k} & ...
& \dsum\limits_{k=0}^{n_{1}}a_{k,1}\left( x\right) \varepsilon _{m}^{k} \\ 
... &  & ... \\ 
\dsum\limits_{k=0}^{n_{m}}a_{k,m}\left( x\right) \varepsilon _{1}^{k} & ...
& \dsum\limits_{k=0}^{n_{m}}a_{k,m}\left( x\right) \varepsilon _{m}^{k}%
\end{array}%
\right\vert ,
\end{equation}%
or by substracting the first column to the following ones

\begin{equation}
S\left( \Phi \mid X\right) =\frac{1}{\Delta (\varepsilon
_{1},...,\varepsilon _{m})}\left\vert 
\begin{array}{cccc}
a_{0,1}\left( x\right) +O(\varepsilon _{1}) & \dsum%
\limits_{k=1}^{n_{1}}a_{k,1}\left( x\right) \left( \varepsilon
_{2}^{k}-\varepsilon _{1}^{k}\right) & ... & \dsum%
\limits_{k=1}^{n_{1}}a_{k,1}\left( x\right) \left( \varepsilon
_{m}^{k}-\varepsilon _{1}^{k}\right) \\ 
... & ... &  & ... \\ 
a_{0,m}\left( x\right) +O(\varepsilon _{1}) & \dsum%
\limits_{k=0}^{n_{m}}a_{k,m}\left( x\right) \left( \varepsilon
_{2}^{k}-\varepsilon _{1}^{k}\right) & ... & \dsum%
\limits_{k=0}^{n_{m}}a_{k,m}\left( x\right) \left( \varepsilon
_{m}^{k}-\varepsilon _{1}^{k}\right)%
\end{array}%
\right\vert ,
\end{equation}%
that is,

\begin{equation}
S\left( \Phi \mid X\right) =\frac{1}{\Delta (\varepsilon
_{2},...,\varepsilon _{m})}\left\vert 
\begin{array}{cccc}
a_{0,1}\left( x\right) +O(\varepsilon _{1}) & \dsum%
\limits_{k=1}^{n_{1}}a_{k,1}\left( x\right) \varepsilon _{2}^{k-1}\left(
1+O(\varepsilon _{1})\right) & ... & \dsum\limits_{k=1}^{n_{1}}a_{k,1}\left(
x\right) \varepsilon _{m}^{k-1}\left( 1+O(\varepsilon _{1})\right) \\ 
... & ... &  & ... \\ 
a_{0,m}\left( x\right) +O(\varepsilon _{1}) & \dsum%
\limits_{k=1}^{n_{m}}a_{k,m}\left( x\right) \varepsilon _{2}^{k-1}\left(
1+O(\varepsilon _{1})\right) & ... & \dsum\limits_{k=1}^{n_{m}}a_{k,m}\left(
x\right) \varepsilon _{m}^{k-1}\left( 1+O(\varepsilon _{1})\right)%
\end{array}%
\right\vert .
\end{equation}

If we take the limit $\varepsilon _{1}\rightarrow 0$, ie $x_{1}\rightarrow x$%
, we obtain

\begin{equation}
\underset{x_{1}\rightarrow x}{\lim }S\left( \Phi \mid X\right) =\frac{1}{%
\Delta (\varepsilon _{2},...,\varepsilon _{m})}\left\vert 
\begin{array}{cccc}
a_{0,1}\left( x\right) & \dsum\limits_{k=1}^{n_{1}}a_{k,1}\left( x\right)
\varepsilon _{2}^{k-1} & ... & \dsum\limits_{k=1}^{n_{1}}a_{k,1}\left(
x\right) \varepsilon _{m}^{k-1} \\ 
... & ... &  & ... \\ 
a_{0,m}\left( x\right) & \dsum\limits_{k=1}^{n_{m}}a_{k,m}\left( x\right)
\varepsilon _{2}^{k-1} & ... & \dsum\limits_{k=1}^{n_{m}}a_{k,m}\left(
x\right) \varepsilon _{m}^{k-1}%
\end{array}%
\right\vert .
\end{equation}

Substracting the second columns to the following ones gives

\begin{eqnarray}
S\left( \Phi \mid x,x_{2},...,x_{m}\right) &=&\frac{1}{\Delta (\varepsilon
_{2},...,\varepsilon _{m})}\left\vert 
\begin{array}{cccc}
a_{0,1}\left( x\right) & a_{1,1}\left( x\right) +O(\varepsilon _{2}) & ... & 
\dsum\limits_{k=2}^{n_{1}}a_{k,1}\left( x\right) \left( \varepsilon
_{m}^{k-1}-\varepsilon _{2}^{k-1}\right) \\ 
... & ... &  & ... \\ 
a_{0,m}\left( x\right) & a_{1,m}\left( x\right) +O(\varepsilon _{2}) & ... & 
\dsum\limits_{k=2}^{n_{m}}a_{k,m}\left( x\right) \left( \varepsilon
_{m}^{k-1}-\varepsilon _{2}^{k-1}\right)%
\end{array}%
\right\vert \\
&=&\frac{1}{\Delta (\varepsilon _{3},...,\varepsilon _{m})}\left\vert 
\begin{array}{cccc}
a_{0,1}\left( x\right) & a_{1,1}\left( x\right) +O(\varepsilon _{2}) & ... & 
\dsum\limits_{k=2}^{n_{1}}a_{k,1}\left( x\right) \varepsilon
_{m}^{k-2}\left( 1+O(\varepsilon _{2})\right) \\ 
... & ... &  & ... \\ 
a_{0,m}\left( x\right) & a_{1,m}\left( x\right) +O(\varepsilon _{2}) & ... & 
\dsum\limits_{k=2}^{n_{m}}a_{k,m}\left( x\right) \varepsilon
_{m}^{k-2}\left( 1+O(\varepsilon _{2})\right)%
\end{array}%
\right\vert  \notag
\end{eqnarray}%
and

\begin{equation}
S\left( \Phi \mid x,x,x_{3},...,x_{m}\right) =\underset{x_{2}\rightarrow x}{%
\lim }S\left( \Phi \mid x,x_{2},...,x_{m}\right) =\frac{1}{\Delta
(\varepsilon _{3},...,\varepsilon _{m})}\left\vert 
\begin{array}{cccc}
a_{0,1}\left( x\right) & a_{1,1}\left( x\right) & ... & \dsum%
\limits_{k=2}^{n_{1}}a_{k,1}\left( x\right) \varepsilon _{m}^{k-2} \\ 
... & ... &  & ... \\ 
a_{0,m}\left( x\right) & a_{1,m}\left( x\right) & ... & \dsum%
\limits_{k=2}^{n_{m}}a_{k,m}\left( x\right) \varepsilon _{m}^{k-2}%
\end{array}%
\right\vert .
\end{equation}

The iteration is immediate and gives the researched result

\begin{equation}
S\left( \Phi \mid x,...,x\right) =\left\vert 
\begin{array}{cccc}
a_{0,1}\left( x\right) & a_{1,1}\left( x\right) & ... & a_{n_{1}-1,1}\left(
x\right) \\ 
... & ... &  & ... \\ 
a_{0,m}\left( x\right) & a_{1,m}\left( x\right) & ... & a_{n_{m}-1,m}\left(
x\right)%
\end{array}%
\right\vert =\frac{W\left( \phi _{1},...,\phi _{m}\mid x\right) }{%
\dprod\limits_{j=1}^{m-1}j!}.
\end{equation}

\section{Confluent limits of the generalized Schur polynomials and of the
generalized Jacobi-Trudi formula}

\subsection{EOP and generalized Schur polynomials}

Sergeev and Veselov \cite{veselov,harnad} have defined the generalized Schur
polynomials associated to the family of polynomials $\left( \Pi
_{n}^{a}\right) $ as

\begin{equation}
S_{\lambda }\left( Z\right) =\frac{\Delta \left( \Pi _{\lambda }\mid
Z\right) }{\Delta \left( \Pi _{0}\mid Z\right) },\ \lambda =(\lambda
_{1},...,\lambda _{m}),
\end{equation}%
where

\begin{equation}
\Delta \left( \Pi _{\lambda }\mid Z\right) =\left\vert 
\begin{array}{ccc}
\Pi _{\lambda _{1}+m-1}^{\alpha }\left( z_{1}\right) & ... & \Pi _{\lambda
_{1}+m-1}^{\alpha }\left( z_{m}\right) \\ 
... &  & ... \\ 
\Pi _{\lambda _{m}}^{\alpha }\left( z_{1}\right) & ... & \Pi _{\lambda
_{m}}^{\alpha }\left( z_{m}\right)%
\end{array}%
\right\vert .
\end{equation}

Note that (see Eq(\ref{Deltamonic}))

\begin{equation}
\Delta \left( \Pi _{0}\mid Z\right) =\Delta \left( Z\right) .
\end{equation}

\bigskip From theorem 1, we then deduce immediately

\begin{equation}
S_{\lambda }\left( z\right) =\underset{\left\{ x_{i}\rightarrow x\right\} }{%
\lim }S_{\lambda }\left( Z\right) =\frac{1}{\dprod\limits_{j=1}^{m-1}j!}%
\mathcal{W}_{\lambda }\left( z\right)   \label{Schur EOP}
\end{equation}%
and the EOP considered above appear to be the confluent limits of
generalized Schur polynomials.

\subsection{Confluent limit of the generalized Jacobi-Trudi formula}

Sergeev and Veselov \cite{veselov,harnad}, have shown that the generalized
Schur polynomials satisfy a generalized Jacobi-Trudi formula of the form

\begin{equation}
S_{\lambda }\left( Z\right) =\left\vert 
\begin{array}{cccc}
S_{\lambda _{1}}^{\left( 0,m\right) }\left( Z\right) & S_{\lambda
_{1}}^{\left( 1,m\right) }\left( Z\right) & ... & S_{\lambda _{1}}^{\left(
m-1,m\right) }\left( Z\right) \\ 
S_{\lambda _{2}-1}^{\left( 0,m\right) }\left( Z\right) & S_{\lambda
_{2}-1}^{\left( 1,m\right) }\left( Z\right) & ... & S_{\lambda
_{2}-1}^{\left( m-1,m\right) }\left( Z\right) \\ 
... & ... &  & ... \\ 
S_{\lambda _{m}-m+1}^{\left( 0,m\right) }\left( Z\right) \mathit{\ } & 
S_{\lambda _{m}-m+1}^{\left( 1,m\right) }\left( Z\right) & ... & S_{\lambda
_{m}-m+1}^{\left( m-1,m\right) }\left( Z\right)%
\end{array}%
\right\vert ,  \label{GJT}
\end{equation}%
where the multivariable polynomials $S_{k}^{\left( i,m\right) }\left(
Z\right) $ verify the recursion relations

\begin{equation}
\left\{ 
\begin{array}{c}
S_{k}^{\left( i+1,m\right) }\left( Z\right) =S_{k+1}^{\left( i,m\right)
}\left( Z\right) +p_{k+m-1,a}S_{k}^{\left( i,m\right) }\left( Z\right)
+q_{k+m-1,a}S_{k-1}^{\left( i,m\right) }\left( Z\right) \\ 
S_{k}^{\left( i+1,m\right) }\left( Z\right) =z_{1}S_{k}^{\left( i,m\right)
}\left( Z\right) +S_{k+1}^{\left( i,m-1\right) }\left( Z\right)%
\end{array}%
\right. ,  \label{recrel}
\end{equation}%
the coefficients $p_{k}$ and $q_{k}$ being given by Eq(\ref{recH}), Eq(\ref%
{recL}) and Eq(\ref{recJ}).

$S_{k\geq 0}^{\left( 0,m\right) }\left( Z\right) $ is the generalized Schur
polynomial associated to a column Young diagram of heigth $k$, that is,%
\begin{equation}
S_{k}^{\left( 0,m\right) }\left( Z\right) =S_{(k,0,...,0)}\left( Z\right) =%
\frac{1}{\Delta \left( Z\right) }\left\vert 
\begin{array}{ccc}
\Pi _{k+m-1}^{a}\left( z_{1}\right)  & ... & \Pi _{k+m-1}^{a}\left(
z_{m}\right)  \\ 
\Pi _{m-2}^{a}\left( z_{1}\right)  & ... & \Pi _{m-2}^{a}\left( z_{m}\right) 
\\ 
... &  & ... \\ 
\Pi _{0}^{a}\left( z_{1}\right)  & ... & \Pi _{0}^{a}\left( z_{m}\right) 
\end{array}%
\right\vert ,
\end{equation}%
which is extended to negative $k$ by setting $S_{k<0}^{\left( 0,m\right)
}\left( Z\right) =0$.

The confluent limit of the generalized JT formula Eq(\ref{GJT})

\begin{equation}
S_{\lambda }\left( z\right) =\left\vert 
\begin{array}{cccc}
S_{\lambda _{1}}^{\left( 0,m\right) }\left( z\right) & S_{\lambda
_{1}}^{\left( 1,m\right) }\left( z\right) & ... & S_{\lambda _{1}}^{\left(
m-1,m\right) }\left( z\right) \\ 
S_{\lambda _{2}-1}^{\left( 0,m\right) }\left( z\right) & S_{\lambda
_{2}-1}^{\left( 1,m\right) }\left( z\right) & ... & S_{\lambda
_{2}-1}^{\left( m-1,m\right) }\left( z\right) \\ 
... & ... &  & ... \\ 
S_{\lambda _{m}-m+1}^{\left( 0,m\right) }\left( z\right) \mathit{\ } & 
S_{\lambda _{m}-m+1}^{\left( 1,m\right) }\left( z\right) & ... & S_{\lambda
_{m}-m+1}^{\left( m-1,m\right) }\left( z\right)%
\end{array}%
\right\vert ,  \label{CGJT1}
\end{equation}%
is a priori different in nature from the Jacobi-Trudi type formula given in
Eq(\ref{NoumiJT}). This last does not appear as the confluent limit of
Sergeev -Veselov's generalized Jacobi-Trudi formula but simply as a
rewriting of Eq(\ref{Schur EOP}) on the basis of the explicit derivation
formula Eq(\ref{derivmonic}). More precisely, in the confluent limit, we
deduce immediately from theorem 1

\begin{eqnarray}
S_{k}^{\left( 0,m\right) }\left( z\right) &=&\frac{1}{\dprod%
\limits_{j=0}^{m-1}j!}W\left( \Pi _{k+m-1}^{a}\left( z\right) ,\Pi
_{m-2}^{a}\left( z\right) ,...,\Pi _{0}^{a}\left( z\right) \mid z\right) \\
&=&\frac{1}{\dprod\limits_{j=0}^{m-1}j!}\left\vert 
\begin{array}{ccccc}
\Pi _{k+m-1}^{a}\left( z\right) & \Pi _{m-2}^{a}\left( z\right) & ... & \Pi
_{1}^{a}\left( z\right) & 1 \\ 
\left( k+m-1\right) \Pi _{k+m-2}^{a_{1}}\left( z\right) &  & ... & 1! & 0 \\ 
... & ... & ... & 0 & 0 \\ 
& \left( m-2\right) ! &  & ... & ... \\ 
\left( k+m-1\right) ...\left( k+1\right) \Pi _{k}^{a_{m-1}}\left( z\right) & 
0 & ... & 0 & 0%
\end{array}%
\right\vert ,  \notag
\end{eqnarray}%
that is, using Eq(\ref{derivmonic})

\begin{equation}
S_{k}^{\left( 0,l\right) }\left( z\right) =\binom{k+l-1}{k}\Pi
_{k}^{a_{l-1}}\left( z\right) =\frac{1}{\left( l-1\right) !}\frac{d^{l-1}}{%
dz^{l-1}}\left( \Pi _{k+l-1}^{a}\left( z\right) \right) .  \label{S0}
\end{equation}

The $\mathit{g}_{k}^{\left( j\right) }\left( z\right) $ functions appearing
in Eq(\ref{NoumiJT}) identify then to the $S_{k}^{\left( 0,l\right) }\left(
z\right) $ rather than to the $S_{k}^{\left( j,l\right) }\left( z\right) $.

The connection between the two Jacobi-Trudy type formulas becomes clearer if
we return to the proof of Eq(\ref{CGJT1}) in the specific confluent case. If
we note%
\begin{equation}
\mathbf{S}_{\lambda }^{\left( i,m\right) }\left( z\right) =\left( 
\begin{array}{c}
S_{\lambda _{1}}^{\left( i,m\right) }\left( z\right) \\ 
... \\ 
S_{\lambda _{m}-m+1}^{\left( i,m\right) }\left( z\right)%
\end{array}%
\right) ,
\end{equation}%
we have

\begin{equation}
S_{\lambda }\left( z\right) =\det \left[ \mathbf{S}_{\lambda }^{\left(
0,m\right) }\left( z\right) ,\mathbf{S}_{\lambda }^{\left( 1,m\right)
}\left( z\right) ,...,\mathbf{S}_{\lambda }^{\left( m-1,m\right) }\left(
z\right) \right] .
\end{equation}

Eq(\ref{recrel}) implies then

\begin{equation}
\mathbf{S}_{\lambda }^{\left( i+1,m\right) }\left( z\right) =z\mathbf{S}%
_{\lambda }^{\left( i,m\right) }\left( z\right) +\mathbf{S}_{\lambda
+1}^{\left( i,m-1\right) }\left( z\right) ,  \label{recS}
\end{equation}%
where $\lambda +1=\left( \lambda _{1}+1,...,\lambda _{m}+1\right) $. By
successively using Eq(\ref{recS}) and combining the columns from the last
one to the second one, we obtain

\begin{eqnarray}
S_{\lambda }\left( z\right) &=&\det \left[ \mathbf{S}_{\lambda }^{\left(
0,m\right) }\left( z\right) ,...,\mathbf{S}_{\lambda }^{\left( m-3,m\right)
}\left( z\right) ,\mathbf{S}_{\lambda }^{\left( m-2,m\right) }\left(
z\right) ,z\mathbf{S}_{\lambda }^{\left( m-2,m\right) }\left( z\right) +%
\mathbf{S}_{\lambda +1}^{\left( m-2,m-1\right) }\left( z\right) \right] \\
&=&\det \left[ \mathbf{S}_{\lambda }^{\left( 0,m\right) }\left( z\right)
,...,\mathbf{S}_{\lambda }^{\left( m-3,m\right) }\left( z\right) ,\mathbf{S}%
_{\lambda }^{\left( m-2,m\right) }\left( z\right) ,\mathbf{S}_{\lambda
+1}^{\left( m-2,m-1\right) }\left( z\right) \right]  \notag \\
&=&\det \left[ \mathbf{S}_{\lambda }^{\left( 0,m\right) }\left( z\right)
,...,\mathbf{S}_{\lambda }^{\left( m-3,m\right) }\left( z\right) ,z\mathbf{S}%
_{\lambda }^{\left( m-3,m\right) }\left( z\right) +\mathbf{S}_{\lambda
+1}^{\left( m-3,m-1\right) }\left( z\right) ,\mathbf{S}_{\lambda +1}^{\left(
m-2,m-1\right) }\left( z\right) \right]  \notag \\
&=&\det \left[ \mathbf{S}_{\lambda }^{\left( 0,m\right) }\left( z\right)
,...,\mathbf{S}_{\lambda }^{\left( m-3,m\right) }\left( z\right) ,\mathbf{S}%
_{\lambda +1}^{\left( m-3,m-1\right) }\left( z\right) ,\mathbf{S}_{\lambda
+1}^{\left( m-2,m-1\right) }\left( z\right) \right]  \notag \\
&=&...  \notag \\
&=&\det \left[ \mathbf{S}_{\lambda }^{\left( 0,m\right) }\left( z\right) ,%
\mathbf{S}_{\lambda +1}^{\left( 0,m-1\right) }\left( z\right) ,\mathbf{S}%
_{\lambda +1}^{\left( 1,m-1\right) }\left( z\right) ,...,\mathbf{S}_{\lambda
+1}^{\left( m-2,m-1\right) }\left( z\right) \right] .  \notag
\end{eqnarray}

Repeating the same procedure on the columns from the last one to the third
one, gives

\begin{equation}
S_{\lambda }\left( z\right) =\det \left[ \mathbf{S}_{\lambda }^{\left(
0,m\right) }\left( z\right) ,\mathbf{S}_{\lambda +1}^{\left( 0,m-1\right)
}\left( z\right) ,\mathbf{S}_{\lambda +2}^{\left( 0,m-2\right) }\left(
z\right) ,\mathbf{S}_{\lambda +2}^{\left( 1,m-2\right) }\left( z\right) ,...,%
\mathbf{S}_{\lambda +2}^{\left( m-3,m-2\right) }\left( z\right) \right] .
\end{equation}

The recursion is immediate and we obtain

\begin{eqnarray}
S_{\lambda }\left( z\right) &=&\det \left[ \mathbf{S}_{\lambda }^{\left(
0,m\right) }\left( z\right) ,\mathbf{S}_{\lambda +1}^{\left( 0,m-1\right)
}\left( z\right) ,\mathbf{S}_{\lambda +2}^{\left( 0,m-2\right) }\left(
z\right) ,...,\mathbf{S}_{\lambda +m-1}^{\left( 0,1\right) }\left( z\right) %
\right] \\
&=&\left\vert 
\begin{array}{cccc}
S_{\lambda _{1}}^{\left( 0,m\right) }\left( z\right) & S_{\lambda
_{1}+1}^{\left( 0,m-1\right) }\left( z\right) & ... & S_{\lambda
_{1}+m-1}^{\left( 0,1\right) }\left( z\right) \\ 
S_{\lambda _{2}-1}^{\left( 0,m\right) }\left( z\right) & S_{\lambda
_{2}}^{\left( 0,m-1\right) }\left( z\right) & ... & S_{\lambda
_{2}+m-2}^{\left( 0,1\right) }\left( z\right) \\ 
... & ... &  & ... \\ 
S_{\lambda _{m}-m+1}^{\left( 0,m\right) }\left( z\right) \mathit{\ } & 
S_{\lambda _{m}-m+2}^{\left( 0,m-1\right) }\left( z\right) & ... & 
S_{\lambda _{m}}^{\left( 0,1\right) }\left( z\right)%
\end{array}%
\right\vert ,  \notag
\end{eqnarray}%
that is,

\begin{equation}
S_{\lambda }\left( z\right) =\left\vert 
\begin{array}{cccc}
\frac{1}{\left( m-1\right) !}\frac{d^{m-1}}{dz^{m-1}}\left( \Pi _{\lambda
_{1}+m-1}^{a}\left( z\right) \right) & \frac{1}{\left( m-2\right) !}\frac{%
d^{m-2}}{dz^{m-2}}\left( \Pi _{\lambda _{1}+m-1}^{a}\left( z\right) \right)
& ... & \Pi _{\lambda _{1}+m-1}^{a}\left( z\right) \\ 
\frac{1}{\left( m-1\right) !}\frac{d^{m-1}}{dz^{m-1}}\left( \Pi _{\lambda
_{2}+m-2}^{a}\left( z\right) \right) & \frac{1}{\left( m-2\right) !}\frac{%
d^{m-2}}{dz^{m-2}}\left( \Pi _{\lambda _{2}+m-2}^{a}\left( z\right) \right)
& ... & \Pi _{\lambda _{2}+m-2}^{a}\left( z\right) \\ 
... & ... &  & ... \\ 
\frac{1}{\left( m-1\right) !}\frac{d^{m-1}}{dz^{m-1}}\left( \Pi _{\lambda
_{m}}^{a}\left( z\right) \right) \mathit{\ } & \frac{1}{\left( m-2\right) !}%
\frac{d^{m-2}}{dz^{m-2}}\left( \Pi _{\lambda _{m}}^{a}\left( z\right) \right)
& ... & \Pi _{\lambda _{m}}^{a}\left( z\right)%
\end{array}%
\right\vert .
\end{equation}

We finally recover Eq(\ref{Schur EOP}) and consequently Eq(\ref{NoumiJT}).

\section{Acknowledgments}

I would like to thank R.\ Milson and D.\ Gomez-Ullate for fruitful exchanges
and helpful suggestions. Thanks also to J.\ Harnad for stimulating
discussions.

\end{document}